# TESERACT: Twin Earth SEnsoR Astrophotonic CubesaT

Authors : Tyler deLoughery, Clayton Lauzon, Haydn Sims
Carleton University, Ottawa, Ontario, Canada
Wahab Almuhtadi
Algonquin College, Ottawa, Ontario, Canada
Ross Cheriton
National Research Council Canada, Ottawa, Ontario, Canada

**Abstract** - In this paper, we evaluate the viability of Cubesats as an attractive platform for lightweight instrumentation by describing a proof of concept CubeSat that houses an astrophotonic chip for transit spectroscopy-based exoplanet atmosphere gas sensing. The Twin Earth SEnsoR Astrophotonic CubesaT (TESERACT) was designed to house a correlation spectroscopy chip along with an electrical and optical system for operation. We investigate design challenges and considerations in incorporating astrophotonic instrumentation such as component integration, thermal management and optical alignment. This work aims to be a pathfinder for demonstrating that astrophotonic-based CubeSat missions can perform leading edge, targeted science in lower-cost CubeSat platforms.

## I. INTRODUCTION

While the cost of launching astronomical payloads to space has decreased over time, a significant financial barrier still remains for smaller organizations[1,2,3,4] to develop and launch space-based astronomical telescopes. As an example, the James Webb Space Telescope was launched December 5th 2021, and had an estimated cost of 10 billion US dollars [5], while projects such as the EUCLID mission were approximately 600 million euros[6]. These costs are largely related to the size and weight of the collecting mirrors. Nevertheless, the instruments add a significant amount of weight and bulk to the to total launch payload

The CubeSat platform follows a standardized framework of various payload configurations offering scientists a cost-effective method to lower the barrier to entry to launch, test and demonstrate novel technologies in space while also serving as an educational platform for students. The CubeSat platform comes in various sizes and resulting orientations. One or multiple "1U" can be outfitted with instrumentation and sent up in orbit. A combination of 1U's may be used for larger payloads or optical systems.

These cost-advantages bring tight restrictions on weight and size of instrumentation payloads that can be placed inside them, leading to reduced functionality, complex integration strategies, and compromises in longevity or performance. In particular, Cubesats which perform optical astronomy face significant challenges in development due to the requirement of an integrated telescope. Astronomical instrumentation such as interferometers and spectrographs typically require long optical system designs with sufficiently large dispersive elements to achieve sensitivity and high spectral resolution, respectively, while maximizing the aperture size to fit the CubeSat smallest dimension. CubeSat platforms could be outfitted with photonic integrated circuits (PIC) to enhance instrument sensitivity while reducing power, weight, physical size, and overall cost[7,8]. The platform provides benefits such as payload miniaturization, weight reduction & lower power consumption resulting in a more condensed & lighter package reducing total cost of launch significantly[7]. Future projects may leverage the relatively low cost launch to implement constellations of CubeSats for spectroscopy or interferometry.

Astrophotonics is the field of guided light-based astronomical instruments, which is a significant departure from the established bulk optics approach to developing instruments, which involves glass elements such as mirrors, lenses, filters, polarizers, beam splitters, etc. Light can be guided using an optical fiber, which acts as a one-dimensional waveguide that can be routed with a high degree of flexibility. Alternatively, light can be routed in two dimensions on a patterned chip in the form of a photonic integrated circuit. Each of these platforms can support light processing by controlling the geometry, material, and refractive index experienced by the light propagating along the optical path in the same way as bulk optics. Astrophotonics brings major advantages in weight and size reduction while being amenable to scalability in a way that cannot be matched by bulk optics.

Recognizing the need to reduce barriers in satellite astronomy and monitoring, the combination of the CubeSat platform with PIC's presents a promising solution for conducting atmospheric science and astronomical science alike[7]. To demonstrate the suitability of PIC's in the CubeSat platform, this paper highlights key findings regarding the advantages and limitations of this approach.

The integration of photonic chips in satellites for optical astronomy has largely not been explored due to the general unsuitability of imaging modes for astrophotonics. The use of astrophotonics in CubeSats has been first investigated using a commercial off-the-shelf lithium niobate ($LiNbO_3$) modulator in a nulling interferometer in the FIRST-S CubeSat project. [9] The optical modulator and integrated optics were used in phasing, amplitude control & beam combination into a single mode fiber. Detection takes place with single photon counting photodiode (SAPD) to reduce readout noise while being able to modulate and demodulate at any frequency. The mission was labeled a "high risk high reward" mission as the precision of the instruments would only allow for the desired effects to take place given the complexity of the system but with the very low cost of the payload, the cubesat offered the ideal platform to conduct this science.

For spectroscopy in particular, astrophotonics has not been pursued yet as traditional bulk optics spectrographs due to the size requirements of dispersive spectrographs. However, integrated photonic spectrographs can obtain very high spectral resolution in a compact chip without the need for a local focal length optical path between mirrors and grating couplers. The two most common types of dispersive spectrographs on a chip include echelle spectrographs and arrayed waveguide gratings which use interference a star coupler and a grating, respectively, to spectrally disperse the light. Our group has developed a correlation spectroscopy technique on a chip using a simple microring resonator tuned to the quasi-periodic absorption line spectrum of a wide range of gas targets. This technique, developed by Vargas et al. [10], uses a bulk optics Fabry-Pérot (FP) cavity, and has shown that multiple absorption lines can be correlated to the FP transmission spectrum with the right mirror spacing. In our case, we develop the correlation filter on a chip, allowing for high levels of multiplexing, low losses, compactness, lightness, high speed modulation of the filter and extreme spectral stability.

We also demonstrate our progress in developing a CubeSat prototype, named Twin Earth SEnsoR Astrophotonic CubesaT (TESERACT), for detection of spectral features from gasses that are expected to be biosignatures in exoplanet atmospheres. The development includes the integration of a silicon-based photonic chip on the 3U CubeSat platform, where an illustration of the CubeSat was shown in Fig 1. We demonstrate the feasibility of PIC's packaged in small payloads such as the CubeSat. The objective is to demonstrate a CubeSat concept that can enable exoplanet transit spectroscopy on targets for extended periods of time to have initial detections of presence of oxygen, carbon dioxide, and methane signals on exoplanets, which will facilitate further follow-up analysis of the planetary characteristics using larger observatories or dedicated satellites such as the James Webb Space Telescope (JWST).

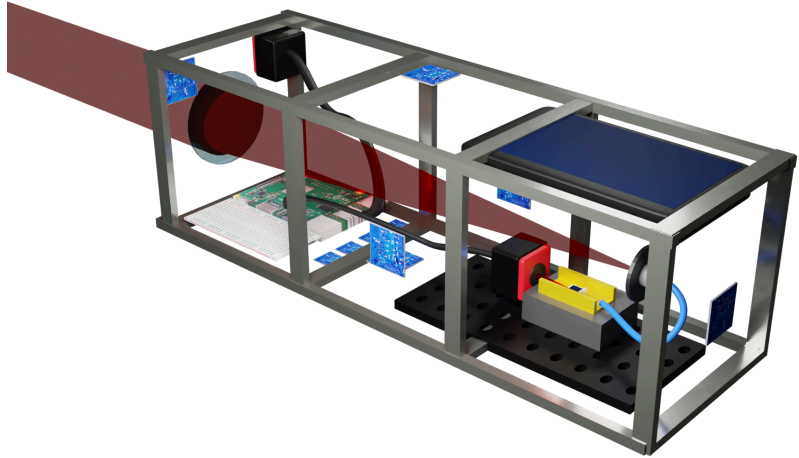

Fig 1. Computer rendered illustration of the TESERACT CubeSat concept.

## II. EXPERIMENTAL SETUP AND RESULTS

*Correlation spectroscopy from exoplanet transit spectroscopy*

For our proof-of-concept CubeSat prototype, our target application is exoplanet atmosphere transit spectroscopy. Given our example of proliferated life on Earth, oxygen, methane and carbon dioxide were selected as our gas targets as biosignatures. From our previous work, we have shown that a microring resonator can be designed as a spectral correlation filter matched to a certain gas, providing reasonable similarity over most of the absorption profile[11]. By modulating the on-chip filter at set frequency, it is possible to achieve gas-specific contrast for many gasses that exhibit quasi-periodic absorption features using synchronous demodulation with a lock-in amplifier. The round trip lengths of the ring have been tuned to generate a transmission filter spectrum with a free spectral range that matches the strongest absorption lines. In the case of $CO_2$, the absorption lines around 1578 nm were targeted for a few reasons. Firstly, the stronger absorption lines at 1435 nm lines are outside the H-band and are strongly blocked by the $H_2O$ lines in that region. Secondly, the lines at 1578 nm are more widely spaced than the ones at the shorter band at ~1572 nm, allowing for improved contrast in the case of pressure and temperature broadening of the spectral lines. Finally, by targeting 1578 nm, we can test cross-sensitivity with the CO band at 1580 nm and 1567 nm. The oxygen band at 765 nm was targeted as it was simply the strongest absorption band that can be targeted. For methane, the lines around ~1650 nm were targeted.

*Photonic integrated circuit design*

The silicon nitride platform is chosen for four reasons: its transparency in the 765 nm oxygen absorption band, low propagation losses, relative compactness, and process accessibility.

A silicon-nitride-on-insulator (SiN on $SiO_2$) strip waveguide (400 nm x 1150 nm) micro-ring resonator in a racetrack geometry was designed with a total round-trip path length of 2.791 mm to match the $CO_2$ absorption spectrum, fabricated by Applied Nanotools[12]. To target the oxygen absorption spectrum, a 400 nm x 400 nm microring resonator was designed to target the 760 nm absorption band. The oxygen absorption spectrum is more complicated than the $CO_2$ absorption spectrum with two combs overlapping the longer wavelength band. Simulations of the correlation performance were undertaken to determine the ideal ring length that maximizes the correlation signal. This round trip ring length is based on the chosen SiN geometry, and was found to be 650 um.

Microheaters made of a tungsten-titanium alloy (TiW) are patterned atop the racetrack rings in order to thermo-optically control the temperature, and hence the wavelength shift of the transmission spectrum ring resonator. The use of Titanium Tungsten (TiW) allows for efficient heaters that are able to be modulated quickly if required. Figure 2A shows the layer structure of the SiN on SiO2 platform. In Figure 2B, a racetrack ring resonator with a tapered fiber input is shown with a through and drop port. The modulated signal expected when gas is present is shown at the drop port output. Figure 2C demonstrates the light path in the waveguide as it travels across the through port back into the 4 channel fiber array. Some light bleed is visible at the left side of the chip due to mode mismatching. Figure 2D shows the light path as it travels from input on the through port and across the ring to the drop port. Transmission graphs are shown on the drop and through port inversely correlated to each other.

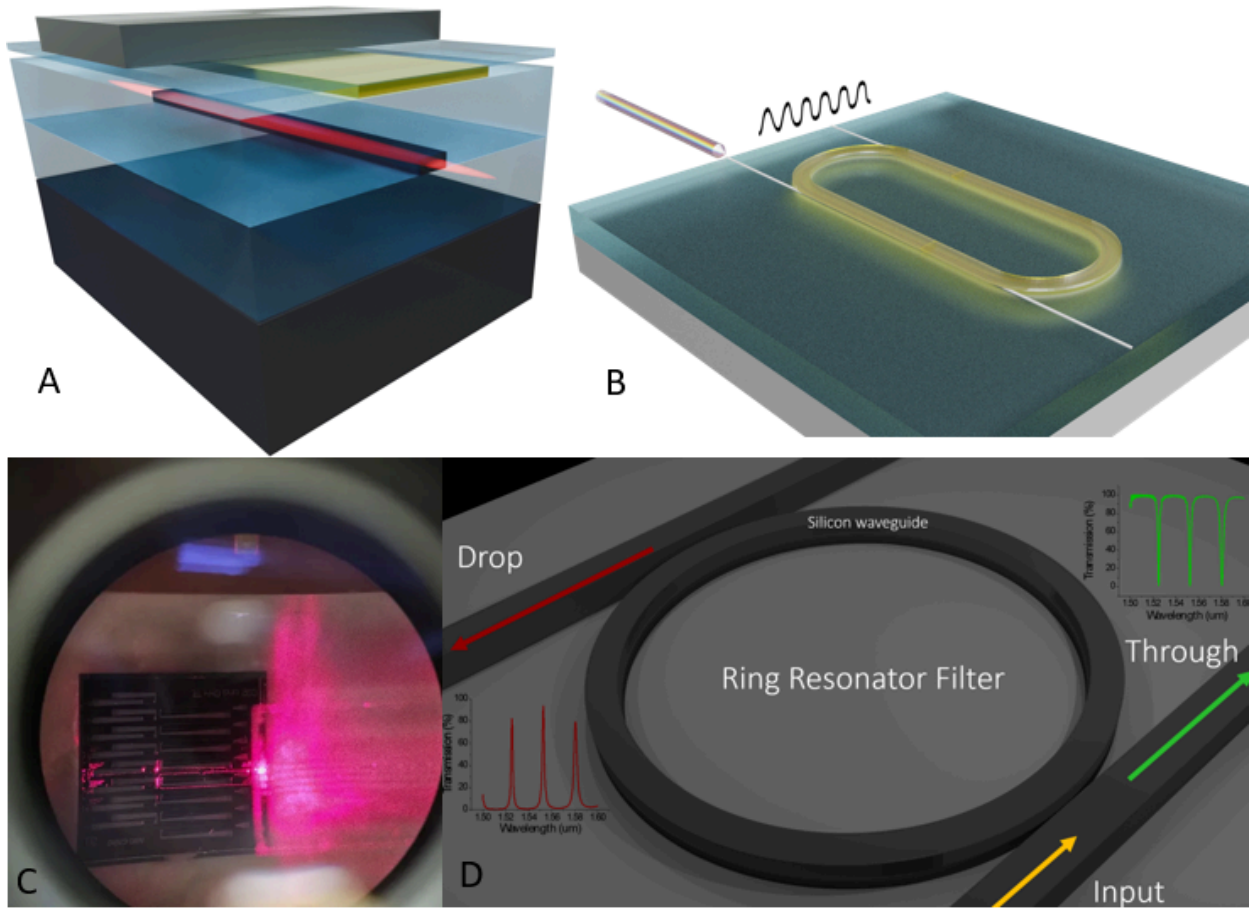

Figure 2. A. Render of Silicon Nitride on Insulator platform, heat traces (gray), heat pads (yellow) glass (blue) waveguide (dark blue). B. Render of ring resonator on chip. C. Ring resonator with visible light injected in the ring (right). D. Ring resonator transmission as seen in the drop and through ports.

*A. Optical System and Photonic Chip*

The 3U CubeSat incorporates collimators, fiber, PIC and cameras as the optical system. A 40 mm aperture on the infrared collimator captures 1580 nm light to detect CO2 and CH4, while a 2cm aperture on the visible collimator captures 765 nm light for the oxygen spectrum.

The optical system follows the pathway as described below:

1. Light enters through a collimating lens.
2. Lens focuses the light into a fiber and directs it toward the photonic chip.
3. Signal travels through a waveguide and across the ring resonator.
4. Signal exits the waveguide back into the fiber.

The PIC serves as the primary instrument onboard, designed to detect and analyze absorption bands of 3 critical gasses being Oxygen (765 nm), Carbon Dioxide (1580 nm) and Methane (1640 nm). The optical system captures incoming light passing by exoplanets, then captured by the optical system onboard the TESSERACT.The signal intensity is analyzed relative to a base signal from the source providing the ability to detect "dips" of intensity in the absorption band of the desired element. The signal is pushed to the chip, filtered and further passed to the photodetector for analysis. The CubeSat incorporates solar cells, a battery, onboard computer systems, cameras, GPS and RF communications outside of the optical system. A Raspberry Pi performs, interprets and transmits that data wirelessly to a ground station computer. Python-based libraries, including NumPy for numerical analysis, Picamera2 for camera control, OpenCV for image processing, and Matplotlib for data visualization, allow for the processing and visualization of these signals.

Due to packaging limitations, only the CO2 ring was able to be applied in the optical system. Using electrically insulating and thermally conductive epoxy, we bonded a 14 pin butterfly package, Thermo Electric Cooler (TEC), Aluminum Niobate thermal pad, and silicon spacers. We further bonded the 4 channel v-groove fiber array to the chip with index matching epoxy. The TEC provides environmental stability throughout the thermal pad and chip attached to a 5 x 30 x 50mm copper plate for heat dissipation. The onboard heaters above the micro ring resonator (MRR) modulate the temperature at a frequency of 1000Hz. To perform the resonant cavity shift required to provide the intensity contrast during analysis, a voltage is applied to the pads using a square wave signal. By integrating the function at varying times during the frequency shift, a transmission contrast can be observed, confirming the presence of the element.

Though the epoxy offers a good bond, it is prone to breakage under high shear stress under sudden tension. Future iteration will include a stronger bond using more precisely placed epoxy and in higher quantities in local surrounding areas to mitigate separation during launch conditions. In this setup, the light from the through port is guided off the chip and into the detection camera, however either the through or drop port can be used. Four fibers, each 1 meter in length are aligned to the waveguide using a 6 axis micropositioner. Due to material and timing constraints, the excess fiber length was coiled rather than spliced to exact lengths according to the specific dimensions and instrument orientations in the CubeSat frame.

The ring resonator is thermally controlled by applying a voltage through a square wave signal to modify the resonant frequency of the ring by approximately 0.1 nm above and below the standard resonant frequency associated with CO2 at 1580 nm. The TiW pads are of 100 μm x 100 μm" in size, while the ring size is expressed by equation 1. The PIC is installed into a 14-pin butterfly package, attached to a copper plate of 25 mm x 55 mm to promote further thermal stability on the chip.

Light throughput was demonstrated at 635 nm, measured at 9.4% of total input at the collimator. Due to mode mismatch between the numerical aperture of the collimator and the use of SMF28e+ fiber, losses are expected with a beam that fills the aperture. Future iterations will include single mode polarization maintaining fibers, independently suited for the visible and infrared spectrum, along with ring resonators with dimensions suited for visible and infrared light.

Losses were measured at 0.2-0.5 dB/cm loss for 1580 nm and 1 dB/cm loss for 765 nm using spiral loss test structures on the same fabrication run.

*Equations*

The equations for microring resonator properties are described below for enabling the on-chip correlation spectroscopy. The first describes the wavelength shift of the resonances via temperature. By increasing the temperature in the waveguide, the effective index of the waveguide changes, leading to a shift in the resonances.

Resonances shift:
$$\Delta \lambda_{TO} = \sigma_{TO} \lambda_0 \left( \frac{\Delta T}{n_g} \right) \quad (1)$$

The second describes the relation of our linewidth spacing inversely correlated with the path length of our ring.

Free spectral range:
$$\text{FSR} = \frac{\lambda^2}{n_g L} \quad (2)$$

$\Delta \lambda_{TO}$ = thermo-optic wavelength shift
$\lambda_0$ = center wavelength
$n_g$ = group index
$\sigma_{TO}$ = thermo-optic coefficient
$\Delta T$ = temperature change
L = ring round trip length

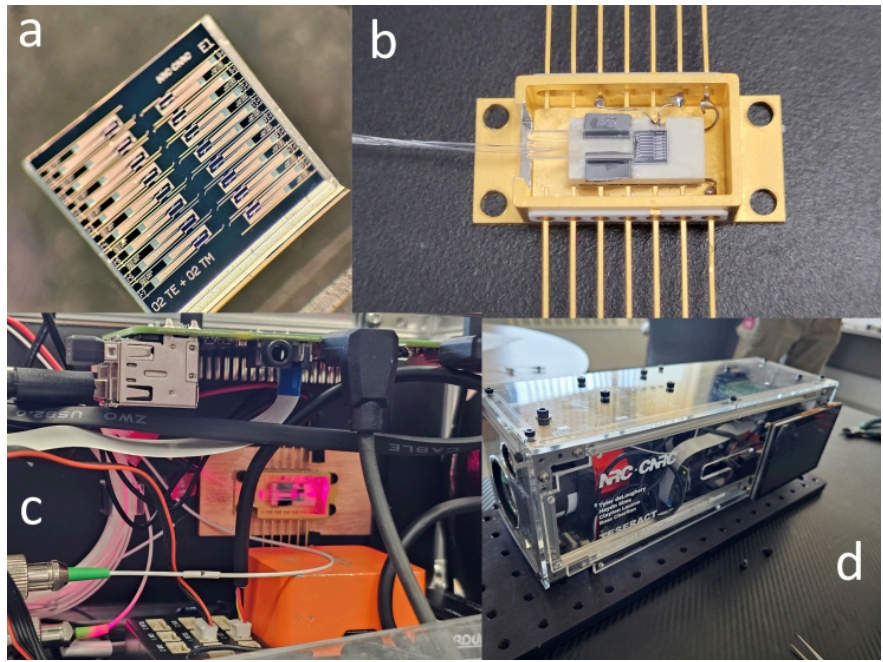

Fig 3. (a) Photo of silicon chip with microring resonator onboard (b) The packaged astrophotonic chip with four channel fiber array (left side) (c) Laser light injected in the optical system (Magenta color) (d) Fully assembled demonstrative 3U CubeSat.

*A. CubeSat design*

Using the 3U Cubesat platform (300 mm x 100mm x 100 mm), it provided the ideal payload size to balance the size of the components along with our strict weight budget of 3 kg (1 kg per 1U)[13]. Given the weak signals involved in this application, we expect apertures at least ~20 cm are required to gather sufficient photons from the exoplanet-star system, however, further study is required to determine the minimum aperture required to resolve a certain absorption depth from a gas spectral line during an exoplanet-star transit. In our case, we projected the optical system was to fit inside a 3U Cubesat frame. A schematic of the optical and electrical system of the CubeSat is shown in Figure 4.

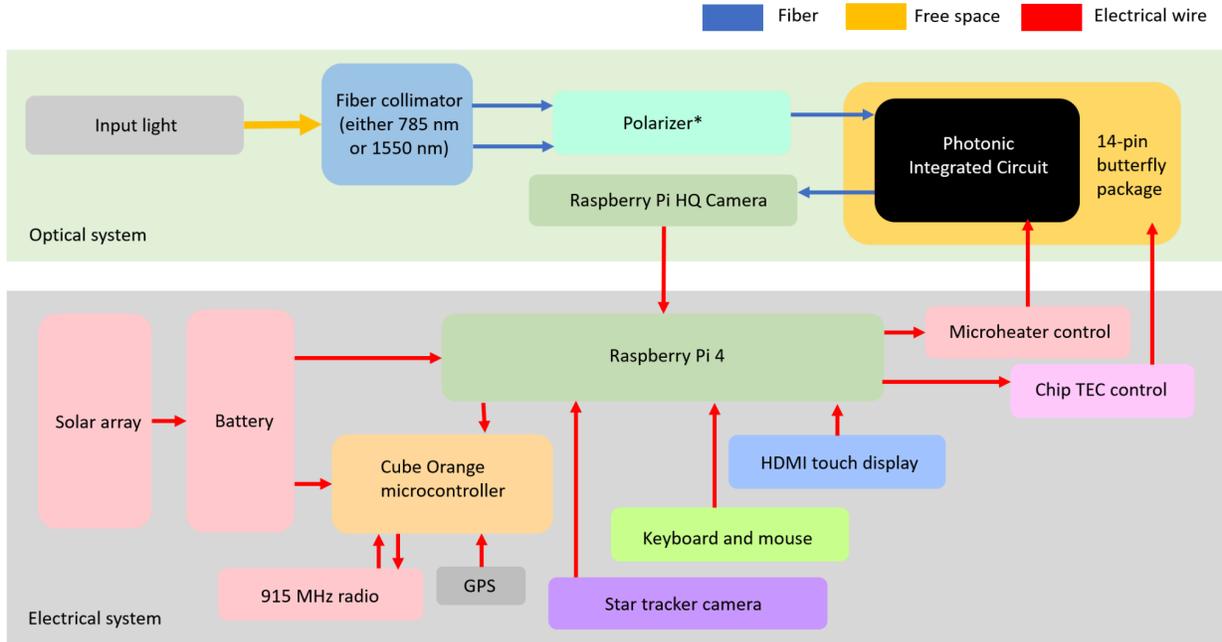

Figure 4. Schematic of the TESERACT optical and electrical system.

*B. PSU and Solar Photovoltaic Charging*

A 40 Ah Solar charged portable power bank was used to power onboard systems. The solar cell is used to maintain power in the battery with a surface area of 20 mm x 60 mm x 140 mm, and provides constant power to charge the battery to power the RaspBerry Pi 4 and Cube Orange microcontroller. The solar cell + battery were successfully installed into the frame of the CubeSat. Solar charging was achieved with the confirmation of changing through a green LED indicator. Facing outwards on the 3U frame, its position ensures it will be facing the Sun at all times both providing Solar charging while simultaneously providing passive solar shielding to the optical and electrical components.

*C. Thermal considerations*

Given the prototype was to be tested on earth, the chip was designed to work at 300 K, given the fluctuations in temperature that occur in LEO, a new chip would need to be designed for those conditions. Thermal shielding was considered, using the battery/solar cell as energy capture while also providing shielding to the components inside. Ideally, the camera system will be kept at the coldest temperature possible, where the CubeSat temperature can drop to about -65 °C with ideal shielding[14]. The fiber and chip-based instrument is less affected by temperature than bulk optics systems, but the temperature of the chip will be set to a certain setpoint to ensure correlation signal stability. While the relative temperature may change by tens of degrees inside the CubeSats with orientation changes, the TEC onboard the optical package provides the necessary temperature control to within ± 0.01 °C to maintain stability of the optical properties on chip. Local Thermal modulation on the chip is driven by the heater control circuit varying the temperature of the chip by ± 10 °C at 1 kHz.

*D. On board telemetry*

In this project, the Cube Orange serves as the microcontroller, equipped with a range of onboard and accessory sensors, including an accelerometer, gyroscope, magnetometer, barometer, RF transmitter, and GPS module. The sensors collect motion, orientation, atmospheric pressure, and location data, wirelessly transmitting to the ground station using the unlicensed ISM 915 MHz radio band. MAVLink UAV protocols were utilized through the DroneKit API. An omnidirectional antenna was employed to ensure a reliable connection regardless of the CubeSat's orientation.

*E. On board processing*

A Raspberry Pi 4 serves as the on-board computer, receiving inputs from two cameras to provide a live feed of the orientation of the satellite along with signal detection from the PIC. Using python scripts, the image is captured every second and converted to grayscale and applies intensity thresholding to ensure faint signals are detected. The center pixel (0,0) is identified, and its intensity is evaluated. If pixel intensity is above this threshold, a "*Light throughput detected*" message is displayed; otherwise, a "No light detected" message is shown. This message is displayed on a window and transmitted to the ground station.

*F. Software*
*a)Ground station:*

The ground station serves as the means of contact with the CubeSat. It's responsible for monitoring the data collected from the sensor suite, the capture window which displays the most recent POV of the PiCam, as well as the forward-facing view of the unit itself. The scripts are written in Python using PyCharm IDE. To minimize development time on the telemetry and astrionics systems, the ground station used the Dronekit API [15], which is an application programming interface that provides methods to communicate and control unmanned aerial vehicles (UAV) and retrieve vehicle state as well as telemetry information. Using the API The onboard flight controller is programmed to update a Tkinter GUI with each GUI label being associated with an attribute or sensor.

*b) On board software:*

A RaspberryPi 4, PiCamera & OpenCV using Python are used to analyze and process images from the Raspberry Pi High Quality Camera. First Capturing an image every second and a half and converting to grayscale. Next any pixels with intensity values greater than or equal to 30 is set to 255 which is the maximum intensity available. Next, finding the center pixel in the image and determining the intensity value at 255, confirming light throughput or 0, proving no light detection.

*G. Material and Weight Budgets*
*a) Weight Budget:*

One of the requirements of the project was to develop a 3U CubeSat with a total weight under 3 kg. As per the 1 kg per 1U standard (3U = 3 kg) [16], the total weight was measured using a portable scale accurate to 0.1 grams. The project managed to achieve a weight of 2.485 kg, as shown in Table 1.

Table 1: Materials Cost Table

| Item | Weight (g) |
|---|---|
| Collimators + mounts | 765 |
| Fiber | 24 |
| Photonic Chip | 5 |
| Cameras | 56 |
| Battery + solar array | 472 |
| CubeOrange | 73 |
| RaspberryPi | 47 |
| GPS + Antenna | 64 |
| Thermoelectric cooler | 34 |
| Wires, tie downs, glues, frame +misc | ~1000 |
| Total | 2485 |

*b) Materials Budget:* The project managed to keep costs below $20,000 CAD. A large portion of the cost was related to the photonic chip costing roughly $12,000 CAD per wafer. However, since the fabrication run produced 36 chips, the cost per chip becomes $333 CAD. The remaining componentry proved to be relatively inexpensive in comparison, using commercial off-the-shelf optomechanical components from Thorlabs, and mechanical and electrical components from various consumer-grade suppliers. For a mission which captures both visible and SWIR light, the instrumentation is expected to cost on the order of $100,000 CAD, mainly due to the SWIR camera for methane and carbon dioxide detection. However, an oxygen-only configuration could be significantly cheaper.

*H. Space Ready Platform Additional Requirements*
*a) Orientation:* A means of orienting the CubeSat will need to be added to allow for the orientation to be controlled. This is required to allow for the CubeSat to point itself at the desired target. A set of 3 magnetorquers or reaction wheels are a common choice for CubeSat orientation to control Pitch, Roll & Yaw.

*b) IR Fiber Camera:* A space qualified SWIR camera with cooling unit to detect extremely small changes in wavelength and intensity in the IR range required for methane and carbon dioxide detection.

*c) Battery and Solar Array*: A space qualified dedicated solar array and battery can replace the current systems to ensure power requirements are met in the vacuum of space under the AM0 solar spectrum.

*d) Sun Sensor:* A Fine tuned digital Sun Sensor (NFSS-411). These sensors permit the CubeSat to know its orientation relative to the sun providing coarse orientation data.

SUMMARY

We show that an optical system can be designed around a packaged astrophotonic chip and fiber-coupled camera. We have developed software to detect and process light into an integrated cubesat telescope, through a photonic chip and into a fiber-coupled camera. This proof of concept demonstrates that astrophotonic instruments can be rapidly developed at low-cost inside cubesat platforms in both the visible and NIR bands. Further development on the prototype CubeSat is required in order to improve throughput, enable oxygen, CO2 and CH4 sensing simultaneously, integrate flight ready components, integrate pointing, and complete the correlation signal processing.